\documentclass[aps,prl,twocolumn,groupedaddress]{revtex4}

\usepackage{graphicx}
\usepackage{dcolumn}
\usepackage{bm}

\begin{document}

\title{Warming in systems with discrete spectrum: spectral diffusion of two dimensional electrons in magnetic field.}

\author{N. Romero Kalmanovitz}
\author {A. A. Bykov}
\altaffiliation{ Institute of Semiconductor Physics, 630090 Novosibirsk, Russia }
\author{Sergey Vitkalov}
\email[Corresponding author: ]{vitkalov@sci.ccny.cuny.edu}
\affiliation{Physics Department, City College of the City University of New York, New York 10031, USA}
\author{A. I. Toropov}
\affiliation{Institute of Semiconductor Physics, 630090 Novosibirsk, Russia}


\begin{abstract}
Warming in complex physical systems, in particular global warming, attracts significant contemporary interest. It is essential, therefore, to understand basic physical mechanisms leading to overheating. It is well known that application of an electric field to conductors heats electric charge carriers. Often an elevated electron temperature describes the result of the heating. This paper demonstrates that an electric field applied to a conductor with discrete electron spectrum produces a non-equilibrium electron distribution, which cannot be described by temperature. Such electron distribution changes dramatically the conductivity of highly mobile two dimensional electrons in a magnetic field, forcing them into a state with a zero differential resistance. Most importantly the results demonstrate that, in general, the effective overheating in the systems with discrete spectrum is significantly stronger than the one in systems with continuous and homogeneous distribution of the energy levels at the same input power.
\end{abstract}

\maketitle

The energy distribution of hot electrons in response to an electric field is a long-standing, interesting problem in condensed matter and plasma physics\cite{levinson}. Despite the simplicity of the experiments, interpretation of results is based significantly on theoretical suggestions. It is widely accepted that overheated, degenerate electron systems are well described by an elevated electron temperature $T_e$. This approach is based on an analysis of the Fokker-Plank type equation with almost elastic scattering \cite{price}. Most dramatically the temperature broadening of the electron distribution function affects kinetic properties of electron systems with a discrete spectrum, in particular, electrons in a magnetic field. Shown by L.D. Landau \cite{landau} a quantization of the electron motion in a magnetic field results in the discrete spectrum of two dimensional electrons. Direct consequences of the Landau quantization are quantum magnetic oscillations in metals:  Shubnikov-de Haas effect and de Haas-van Alphen effect \cite{shoenberg,ando}. At the thermal equilibrium the amplitude of the quantum oscillations depends exponentially on the electron temperature $T_e$. Often the oscillations are used as an electron thermometer. 
 
\begin{figure}
\includegraphics{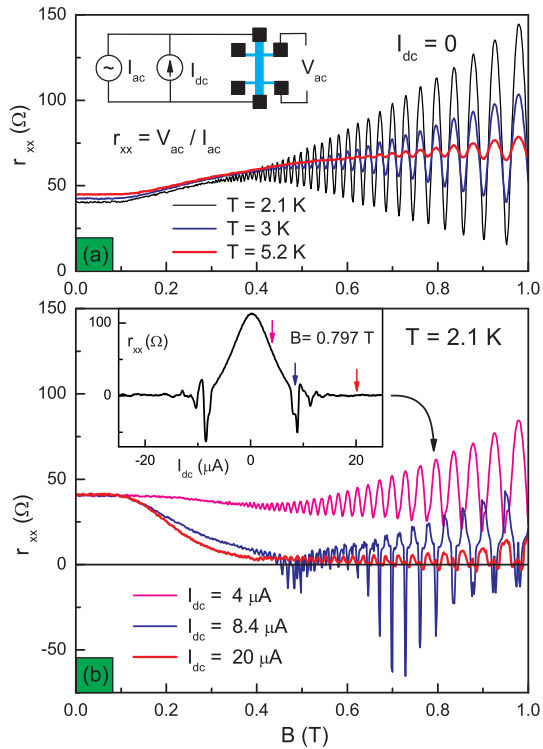}
\caption{(Color online) (a) Dependence of longitudinal resistance $r_{xx}$ on magnetic field B with no $dc$ bias $I_{dc}$ applied at different temperatures as labeled. Insert presents the experimental setup. (b) Dependence of longitudinal resistance $r_{xx}$ on magnetic field B at different $dc$ bias $I_{dc}$ as labeled. Insert presents the bias dependence at fixed magnetic field $B$=0.797 (T). Temperature is 2.1 (K). 
 \label{fig.1}   }
\end{figure}

Below we show that temperature $T_e$ is not an appropriate parameter for   2D electrons overheated by a direct current (DC) in magnetic field. In other words the non-equilibrium distribution function of the DC biased 2D electrons is not described by a temperature. The DC bias originates a non-uniform spectral diffusion of the 2D electrons over the discrete Landau spectrum resulting in a non-equilibrium electron distribution oscillating with the energy.  Our finding is in agreement with recent theoretical consideration of the effect of the DC electric field on the conductivity of 2D electrons \cite{dmitriev} and recent experiments in this field \cite{bykov1,zudov1,bykov2,zudov2}.

Our samples are cleaved from a wafer of a high-mobility GaAs quantum well
grown by molecular beam epitaxy on semi-insulating (001) GaAs
substrates. The width of the GaAs quantum well is 13 nm. 
Three samples (N1,N2,N3) are studied with electron density $n_1$=8.2 $\times
10^{15}$ (m$^{-2}$), $n_2$=8.4$\times 10^{15}$ (m$^{-2}$), $n_3$=8.1$\times 10^{15}$ (m$^{-2}$)  and mobility $\mu_1$=85 (m$^2$/Vs), $\mu_2$=70 (m$^2$/Vs) and $\mu_3$=82 (m$^2$/Vs) at T=2K.
Measurements are carried out between T=0.3K and T=10 K in magnetic
field up to 1.5 T  on $d$=50 $\mu m$ wide Hall bars with a distance of 250
$\mu m$ between potential contacts. The differential longitudinal resistance is
measured at a frequency of 77 Hz in the linear regime.  Direct electric current (dc bias) was applied simultaneously with ac excitation through the same current leads (see insert to Fig. 1).  All samples demonstrate similar behavior. Below we present data for sample N1.

Figure 1a shows the longitudinal resistance $r_{xx}=dV_{xx}/dI$ \cite{resistance} as a function of the magnetic field $B$ at zero DC bias and at different temperatures as labeled. At $B>$0.4 (T) the resistance exhibits the Shubnikov de Haas oscillations. The amplitude of the oscillations decreases significantly with temperature increase. The temperature reduction of the SdH amplitude is result of an effective (exponential) cancellation of the periodic oscillations of the electron conductivity $\sigma(\epsilon)$ with energy $\epsilon$ weighted with the first derivative of the equilibrium (Fermi-Dirac) distribution function $df/d\epsilon$ and averaged over all possible energies \cite{ando}:

$$
    \sigma= \int \sigma(\epsilon)(-d f/d\epsilon) d\epsilon \label{sigma} \eqno(1)
$$
   
Figure 1b shows the longitudinal resistance $r_{xx}=dV_{xx}/dI$ as a function of magnetic field $B$ taken at three different $dc$ biases as labeled at a lattice temperature T=2.1K. The $dc$ electric field changes significantly the longitudinal resistance in magnetic fields above $0.15 (T)$. At $B>0.15$ (T) the disorder broadening $\hbar/ \tau_q$ of the Landau levels is less than the Landau level separation $\hbar \omega_c$ and the electron energy spectrum (density of states) becomes modulated. The modulation increases with the magnetic field. At $B \sim 1$ (T) the Landau levels are completely separated from each other and the electron spectrum is discrete. 

The insert to fig.\ref{fig.1}b demonstrates a typical dependence of the longitudinal resistance $r_{xx}$ on the $dc$ bias in a magnetic field ($B=0.797$ (T)), corresponding to an oscillation maximum. For small dc bias the resistance decreases approaching zero. As the dc bias is raised further, the resistance exhibits a reproducible sharp negative spike at $I_{th}$=8.4 ($\mu$A) and, then, stabilizes near zero at higher biases. The zero differential resistance state is observed in a broad range of magnetic fields at temperatures below few Kelvins\cite{bykov2,zudov2}. 

\begin{figure}
\includegraphics{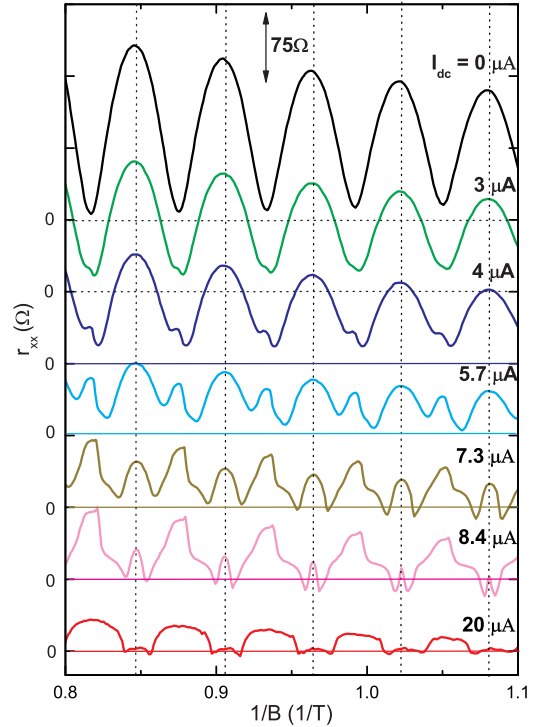}
\caption{ (Color online). Longitudinal resistance $r_{xx}$ plotted vs inverse magnetic field 1/B at different $dc$ bias as labeled. Temperature T=2.1 (K). 
 \label{fig.2}   }
\end{figure}

\begin{figure}
\includegraphics{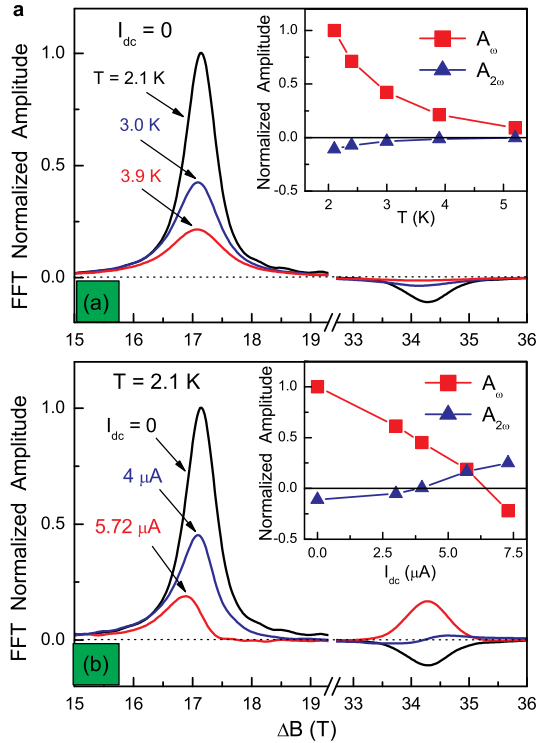}
\caption{ (Color online). (a) Fourier spectrum of the quantum oscillations at different temperature as labeled. Insert shows temperature dependence of amplitude of first (at 17.1 T) and second (at 34.2 T) harmonics of the quantum oscillations. Zero $dc$ bias. (b) Fourier spectrum of the quantum oscillations at different $dc$ bias as labeled. Insert shows the bias dependence of amplitude of first (at 17.1 T) and second (at 34.2 T) harmonics of the quantum oscillations. Lattice temperature is 2.1K.    
 \label{fig.3}   }
\end{figure}

The curves presented in fig.\ref{fig.1}b reveal a dramatic and, in many respects, unexpected evolution of the SdH oscillations with the $dc$ bias. The behavior is significantly different from the temperature evolution shown in fig.1a. Both oscillating and smooth parts of the dependences in fig.1b exhibit substantial changes. The smooth part of the resistance displays a strong reduction with the $dc$ bias in magnetic fields $B>0.15$ (T), at which the density of electron states oscillates with energy. The oscillating part of the resistance reveals a rather complicated behavior with the $dc$ bias.

Figure 2 presents the bias induced evolution of the quantum oscillations in greater detail. Plotted as a function of the inverse magnetic field $1/B$ the oscillations are periodic \cite{shoenberg}. Although the fundamental periodicity is the same for all biases, the form of the oscillations and/or the harmonic content are changed significantly with the bias. The most dramatic changes occur at maxima of the oscillations. With the increasing bias the maxima evolve gradually into minima with essentially zero differential resistance. On the other hand the minima tend to form maxima and, finally, separate the zero resistance regions at the end of the evolution.  For further discussion it is important to realize that the high frequency content of the oscillations with respect to the fundamental periodicity increases with the $dc$ bias increase.

Figure 3a demonstrates quantitatively the evolution of the first and second harmonics of the quantum oscillations with temperature at zero bias. At thermal equilibrium the distribution function $f$ does not oscillate with energy and, therefore, the harmonic content of the quantum oscillations is determined $only$ by the harmonic content of the conductivity $\sigma(\epsilon)$ oscillating  with energy $\epsilon$ due to Landau quantization. The amplitudes of the first $A_{\omega}(T)$ and the second $A_{2\omega}(T)$ harmonics exhibit exponential decrease with the temperature due to the spectral average of the oscillations of the conductivity $\sigma(\epsilon)$ over the range of energies $\Delta \epsilon$ corresponding to the thermal broadening of the distribution function  $\Delta \epsilon \sim kT $. At higher frequency the spectral cancellation is more effective and, as a result, the higher harmonics of the quantum oscillations ($A_{2\omega}(T)$) are exponentially smaller than the amplitude of the fundamental periodicity $A_{\omega}(T)$. The insert to fig.\ref{fig.3}a demonstrates that the amplitudes of the first $A_{\omega}$ and the second  $A_{2\omega}$ harmonics as well as their ratio $R= A_{2\omega}/ A_{\omega}$  decrease significantly with the temperature and approach the zero value asymptotically.

Figure 3b demonstrates the effect of the $dc$ bias on the spectrum of the quantum oscillations. Notably different behavior of the first and, especially,  the second harmonics with $dc$ bias is observed.  The first (fundamental) harmonic does not approach the zero value asymptotically. Instead, as it is shown on the insert to fig.3b, the amplitude crosses the zero value and becomes negative at $I_{dc}=7.5$ $\mu$A. In other words, the phase of the fundamental oscillations changes by 180$^0$.  This corresponds to the evolution presented in figure 2b, where maxima become minima with the bias increase. Even more unusual is the behavior of the second harmonic of the oscillations. One can see that the magnitude of second harmonics does not exponentially decrease with the $dc$ bias contrary to the usual expectation. Instead the second harmonic crosses the zero and reaches a value, which is even bigger than the amplitude $ A_{2\omega}(0)$ at zero bias. At the same time the ratio $R$ between the amplitudes of the second and first harmonics is on the order of unity, which, again, cannot be explained by the temperature broadening of the distribution function. Thus the results indicate that the effect of the direct current on resistance of the 2D electron system with discrete energy spectrum cannot be described by an increase of the electron temperature.

The enhancement of the higher harmonics of the quantum oscillations with the DC bias follows directly from eq.(1) assuming additional, bias induced $oscillations$ of the non-equilibrium distribution function with the energy.  Such oscillations have been found theoretically in DC biased 2D electron systems recently\cite{dmitriev}. Due to the conservation of total electron energy in the presence of an electric field $\vec E$ and elastic electron-impurity scattering the kinetic energy of an electron depends on electron position $\vec r$: $\epsilon(\vec r)=\epsilon_0+e \vec E \vec r$. As a result, diffusion motion of the electron in real space originates a diffusion of the electron in energy space. The coefficient of the spectral diffusion $D_\epsilon$ is proportional to the coefficient of the spatial diffusion D: $D_\epsilon=(e E)^2 D \sim (\delta \vec r)^2$. The spatial (and so spectral) diffusion is most effective in the center of the Landau levels, where the density of states is high, gradually decreases away from the center and is suppressed considerably between Landau levels, where the density of states is small.  

The spectral diffusion generates an electron spectral flow $J_\epsilon$ from low energy regions (occupied levels) to high energies (empty levels). The spectral flow is proportional to $D_\epsilon$ and to the gradient of the distribution function $df/d\epsilon$: $J_\epsilon =D(\epsilon) \cdot df/d\epsilon$. In a stationary state the spectral electron flow $J_\epsilon$ is constant. As a result, the gradient of the distribution function $df/d\epsilon$ is strong in the regions of weak spectral diffusion (between Landau levels) and is small in the regions with strong spectral diffusion (centers of the Landau levels). 

A weak inelastic relaxation of the distribution function with an inelastic scattering rate 1/$\tau_{in}$ exists at low temperature\cite{dmitriev}. The weak inelastic scattering does not change considerably the spectral diffusion. However on long times $t \gg \tau_{in}$ the inelastic processes stabilize the bias induced deviations of the distribution function from the thermal equilibrium.  It is taken into account below in a numerical simulation of the spectral diffusion. The spectral diffusion equation (see eq. (9) in \cite{dmitriev}) is solved numerically.

\begin{figure}
\includegraphics{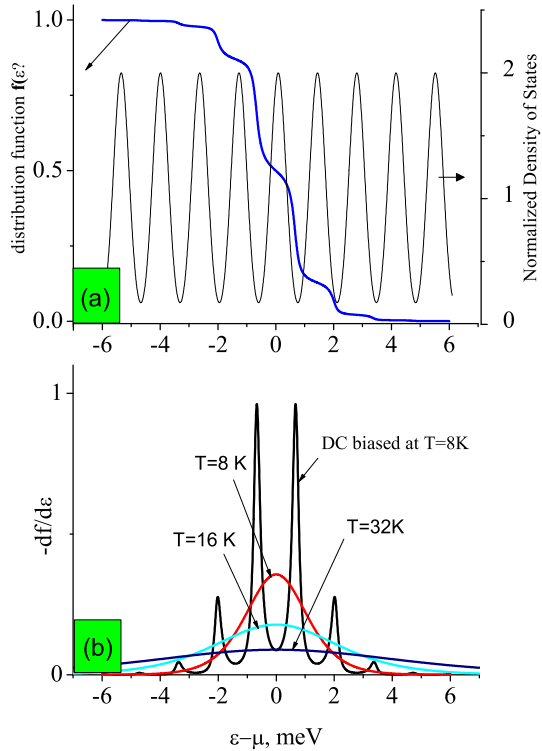}
\caption{ (Color online). (a) right axes: density of states of 2D electrons in magnetic field B=0.797 (T) vs energy; left axes: numerical calculations of electron distribution function modified by the spectral diffusion. The following parameters are used in the simulation: magnetic field B=0.797 T, temperature T=8 K, $dc$ electric field $E=1003$ (V/m), disorder broadening of Landau levels $\hbar / \tau_q$=$\hbar/1.9$ (1/ps) and inelastic scattering time $\tau_{in}=150$ (ps). The times are obtained by methods presented in paper [7]; (b) gradient $df/d\epsilon$ of the distribution function modified by the spectral diffusion shown above and the gradient of the thermal equilibrium function corresponding to different temperatures as labeled.    
\label{fig.4}   }
\end{figure}

Results of the numerical simulation of the spectral diffusion are shown in figure 4a. As expected the spectral diffusion produces periodic variations of the distribution function $f(\epsilon)$ with energy. It is clear, that an elevated temperature $T_e$ cannot describe this oscillating behavior. The additional oscillations of the distribution function enrich significantly the harmonic content of the product $\sigma(\epsilon)\cdot (-df/d\epsilon)$ and, therefore, in accordance with eq.(1), the spectrum of the quantum oscillations. Furthermore shown in fig.4b the considerable decrease of the value $df/d\epsilon$ at the maximums of the conductivity $\sigma(\epsilon)$ and/or density of states makes the net conductivity $\sigma$ (see eq.1) significantly smaller than the unbiased value. This explains the strong reduction of the conductivity (and resistivity $r_{xx}$ \cite{resistance}) with the DC bias at magnetic field $B>0.15$ T shown in fig.1b. 

The strong reduction of the gradient of the electron distribution function $df(\epsilon)/d\epsilon$ near the maximums of the conductivity $\sigma(\epsilon)$ and/or the density of states imitates a strong overheating in these energy regions. Indeed fig.4b demonstrates that at a maximum of the density of states located at the Fermi level $\mu$: $\epsilon-\mu=0$, the gradient of the DC biased distribution function $df(\epsilon)/d\epsilon$ equals the one corresponding to the thermal equilibrium at temperature of 32 K. Thus at the conductivity maximum $\sigma(\epsilon=\mu)$ the effective overheating of the 2D electrons (32K) is significantly stronger the actual additional broadening of the distribution function ($\sim 1-2K$) induced by the DC bias. This provides the dramatic change of the transport properties of the system (see fig.1b). 

The strong reduction of the $df/d\epsilon$ is result not only of the high spectral diffusion near the spectral maxima, but, mostly, is a result of the $weak$ spectral diffusion between them. In fact the weak diffusion between the spectral maxima controls the global spectral flow $J_\epsilon$. In some sense the regions of the phase space with the weak spectral diffusion provide a "thermal" isolation of the regions with the strong spectral diffusion. This DC bias induced, thermal stratification of the phase space facilitates the strong overheating in the regions essential for the net electron transport.
Similar spectral diffusion and the thermal stratification are expected in 2D electron systems affected by an electromagnetic radiation. In this case the spectral diffusion is a result of quantum transitions between energy levels induced by the external radiation. This points toward a generic nature of the phenomena. Thus the presented results suggest, that in systems with a discrete or strongly modulated energy spectrum the effective overheating is significantly $stronger$ than the one in systems with weakly modulated, homogeneous spectrum at the same input power.

\begin{acknowledgments}

This work was supported by NSF: DMR 0349049 and RFBR, project
No.08-02-01051 

\end{acknowledgments}


\begin{references}
\bibitem{resistance}In strong magnetic field the longitudinal resistance $r_{xx}$ is proportional to longitudinal conductivity $\sigma$ \cite{dmitriev}. 
In this paper we consider these two terms to be similar.
\end{references}

\end{document}